\documentclass[12pt,a4paper]{article}

\usepackage[T1]{fontenc}

\usepackage{lmodern}
\usepackage[utf8]{inputenc}
\usepackage{cite}
\usepackage{hyperref}
\usepackage{setspace}
\hypersetup{
	colorlinks=true,
	linkcolor=dark-blue,
	citecolor=dark-red,
	urlcolor=dark-green,
	linktoc=page
}
\usepackage{color}
\definecolor{dark-gray}{gray}{0.20}
\definecolor{gray}{gray}{0.30}
\definecolor{light-gray}{gray}{0.80}
\definecolor{dark-red}{rgb}{0.7,0,0}
\definecolor{dark-green}{rgb}{0.1,0.4,0}
\definecolor{dark-blue}{rgb}{0.3,0.3,0.7}
\definecolor{light-blue}{rgb}{0.8,0.8,1}
  \usepackage{a4wide}
  \usepackage{latexsym}
  \usepackage{epsf}
  \usepackage{bbm}
  \usepackage{graphicx}
  \usepackage{amsmath,amssymb,amsthm}
\usepackage{verbatim}

\renewcommand{\d}{\textrm{d}}

\newcommand{\f}[2]{\frac{#1}{#2}}
\newcommand{\e}{\textrm{e}}

\newcommand{\be}{\begin{equation}}
\newcommand{\ee}{\end{equation}}
\newcommand{\vol}{\text{vol}}
\newcommand{\Tthree}[0]{\mathbb{T}^3}

\csname @addtoreset\endcsname{equation}{section}

\usepackage{tikz}
\usetikzlibrary{arrows,automata,positioning,calc,trees,decorations.pathmorphing,decorations.markings,3d,fadings}
\tikzset{snake it/.style={decorate, decoration={snake,segment length=3pt, amplitude=1pt}}}
\usepackage{tikz}
\usetikzlibrary{positioning,arrows}
\usetikzlibrary{decorations.pathmorphing}
\usetikzlibrary{decorations.markings}
\usetikzlibrary{decorations.text}
\usetikzlibrary{calc}

\usetikzlibrary{arrows,automata,positioning,calc,trees,decorations.pathmorphing,decorations.markings,3d}
\usepackage{pgfplots}

\usepackage{tikz}

\usetikzlibrary{decorations.markings}

\begin{document}

\begin{center}

{\LARGE {\bf Smearing orientifolds in flux \\ \vspace{0.3cm}compactifications can be OK}}  \\

%\\ \vspace{0.4cm} in massive IIA supergravity}

\vspace{1.5 cm} {\large  Stephanie Baines$^{a}$ and Thomas Van Riet$^b$  }\\

\vspace{.5 cm} ${}^a$ Theoretical Particle Physics and Cosmology Group, Department of Physics,\\ King’s College London, Strand, London WC2R 2LS, UK . \\
${}^b$ Instituut voor Theoretische Fysica, KU Leuven,\\
Celestijnenlaan 200D B-3001 Leuven, Belgium.\\

\vspace{0.7cm} {\small \upshape\ttfamily  
\href{mailto:stephanie.baines@kcl.ac.uk}{stephanie.baines@kcl.ac.uk} , \href{mailto:thomas.vanriet@kuleuven.be}{thomas.vanriet@kuleuven.be}}  \\

\vspace{2cm}

{\bf Abstract}

\end{center}

{\small  We present explicit examples of supergravity solutions corresponding to backreacting localised (non-intersecting) O6 planes in flux reductions of massive IIA supergravity and address some criticism towards the very existence of such solutions. We verify in detail how the smeared orientifold solution becomes a good approximation to the localised solution in the large volume/weak coupling limit, as expected. We also find an exotic solution where prior to backreaction the internal space has a boundary and when backreaction is included the boundary disappears and the space closes off. The exotic example is however outside of the supergravity approximation everywhere. %in contrast with the $\mathbb{T}^3/\mathbb{Z}_2$ flux solution.  }

\newpage

%\tableofcontents
%\newpage
\section{Introduction}

It is well known that moduli stabilisation in flux reductions seems possible in massive IIA string theory without invoking quantum or derivative corrections to 10D supergravity if orientifolds are present \cite{Derendinger:2004jn, DeWolfe:2005uu}. Using the general structure of 4D $N=1$ supergravity theories obtained from IIA orientifolds developed in \cite{Grimm:2004ua}, it was found that Calabi-Yau O6 models can indeed achieve full moduli stabilisation at the classical level \cite{Derendinger:2004jn, DeWolfe:2005uu}. More striking is the property that the AdS-length scale of these vacua can be made arbitrarily large with respect to the KK scale by cranking up the $F_4$ flux which is unbounded by tadpoles. Futhermore, in this same limit, the string coupling goes to zero and all length scales and inverse curvature scales go to infinity suggesting that the solution is under arbitrary control from corrections. To our knowledge, this is the only 4D flux solution that achieves all this magic together: full moduli stabilisation, scale separation, and arbitrary weak coupling and curvatures. The ``T-duality" inspired sisters in IIB constructed in \cite{Caviezel:2009tu, Petrini:2013ika} are a bit less understood in our opinion. Recently moduli stabilised AdS$_3$ with arbitrary scale separation were constructed from G2 compactifications with intersecting O6 planes and Romans mass \cite{Farakos:2020phe}. 

Despite all the nice properties of these flux solutions, doubts about their stringy consistency \cite{Banks:2006hg, McOrist:2012yc} gave rise to a ``Swampland conjecture'' which suggests that scale separation is not allowed in consistent string theory models \cite{Lust:2019zwm} (see also \cite{Gautason:2018gln, Blumenhagen:2019vgj} for closely related statements). Setting aside these worries surrounding the IIA vacua, the AdS Swampland conjectures also have roots in a generalisation of the distance conjecture \cite{Ooguri:2006in},\footnote{See \cite{Brennan:2017rbf, Palti:2019pca} for an overview of the coherence and circumstantial evidences for the various Swampland conjectures. See also \cite{Font:2019uva, Lust:2020npd} for the latest investigation on the matter of scale separation.} the no-go for scale separation in flux compactifications without sources \cite{Gautason:2015tig} and even in the rather unfamiliar properties of the would-be dual CFT$_3$ \cite{Aharony:2008wz, Polchinski:2009ch, Alday:2019qrf}. 

The inspiration for this note arose while considering the effects of backreaction from O6 planes in massive IIA supergravity \cite{Banks:2006hg, McOrist:2012yc}. As pointed out in \cite{Acharya:2006ne}, the 4D supergravity solutions of \cite{DeWolfe:2005uu} solve the 10D equations with \emph{smeared} O6 plane sources, whereas the proper backreaction will certainly bring us away from the CY limit, all the way to a full blown $SU(3)\times SU(3)$ structure as shown in \cite{Saracco:2012wc}. Whether the solution with proper backreaction still obeys the nice features described in \cite{DeWolfe:2005uu} is therefore up for debate. On the other hand, the smearing of sources is rather well understood in IIB no-scale vacua and T-duals thereof \cite{Grana:2006kf, Blaback:2010sj}, where the main effect is the appearance of a warp factor in front of the orientifold worldvolume directions and an inverse factor along the transverse space. Although the solution is altered, it does not disappear and the complex-structure moduli (or T-duals of them) remain in the same position. That said, we remain open minded to the possibility that the backreaction of the O6 planes in scale separated AdS vacua is problematic. Recently two papers \cite{Junghans:2020acz,Marchesano:2020qvg} made progress in understanding these matters. Reference \cite{Junghans:2020acz} moves towards localisation by adding a correction to the smeared solution at first order in the inverse $F_4$-flux and finds no obstacles in analysing the 10D equations of motion. Of course the solution is corrected away from the Calabi-Yau regime, but the nice features are not ruined in that first-order approximation. Similarly, reference \cite{Marchesano:2020qvg} analyses the 10D killing spinor equations together with the Bianchi identities and finds similar results. These papers describe the progression away from the CY limit towards an $SU(3)\times SU(3)$ structure manifold, but do not settle the issue completely for two reasons. Firstly, we do not know whether the 10D solution remains consistent at higher orders. This might not be the case since the higher orders are sensitive to the intersection of the O6 planes. Secondly, even if a fully localised supergravity solution with intersecting sources can be found, or argued to exist, there is no clear understanding of the nature of such O6 singularities due to a lack of an M-theory lift in the presence of a Romans mass. 

In this paper we focus on models in which the localisation of the O6 plane in massive IIA is understood completely in the sense that there is an ansatz which solves the 10D equations with localised sources. The specific model was studied in detail in \cite{Blaback:2013taa, Blaback:2015gwa, Blaback:2019ucp}. It describes a flux compactification of massive IIA with space-filling O6 planes down to a SUSY-breaking, 7D Minkowski space with flat moduli directions at the classical level. While this solution is not a scale separated AdS vacuum, it shares at least the similar ingredients.\footnote{But not the crucial ingredient of \emph{intersecting} O6 planes.} In particular, the solution can be tuned to arbitrary weak coupling and large volume so that SUGRA remains a fine approximation away from the sources. The reason is that a linear combination of the dilaton and volume is massless and the only constraint is that their product is fixed by flux quanta $K, M$:
\begin{equation}\label{KM}
g_s Vol =(2\pi)^3\frac{K}{M}\,. 
\end{equation}
So nothing forbids, at the classical level, taking $g_S\rightarrow 0$ and $Vol\rightarrow \infty$. As the solution is Minkowski at the classical level, it is scale separated. But this notion of scale separation is not useful without full moduli stabilisation, neither does it come into tension with any Swampland bound. This background also seems to refute certain reservations about O6 planes in Romans supergravity, at least those related to solving 10D equations. But we remain agnostic about deeper lying string theoretic issues. 

Another main point of this note is to show a surprising aspect about these solutions. In particular they come in two classes. In the conventional class, the internal manifold is $\mathbb{T}^3/\mathbb{Z}_2$ and proper backreaction of the O6 planes only dresses the solution with warp factors. The main features such as equation (\ref{KM}) remain unaltered. Appart from the region near the O6 planes, the backreacted solution is weakly coupled and curved. A second class of solutions exists which defeats standard EFT reasoning. Prior to backreaction these solutions are problematic since the internal space has a boundary. Rather remarkably, including backreaction closes off the boundary into a well behaved O6 singularity. However crucial aspects such as equation (\ref{KM}) do not apply and there is no flat direction. This violation of the EFT can be understood from the presence of the boundary in absence of proper backreaction as we explain.

\section{The 7D EFT for dilaton and volume}
Consider massive type IIA supergravity with O6 planes that fill 7 non-compact dimensions and that are point-like inside the 3 remaining compact dimensions. There is only an $H_3$-flux filling the compact dimensions and a Romans mass $F_0$. The kinetic term and effective potential for the dilaton and volume fields in this compactification were computed in \cite{Blaback:2015gwa}  and we recall them here. The ansatz in the 10D Einstein frame is
\begin{equation}
ds^2 = e^{2\alpha\varphi}ds_7^2 + e^{2\beta\varphi} ds_3^2\,,
\end{equation}
where $\alpha= \frac{1}{4}\sqrt{\frac{3}{5}}$ and $\beta=-\frac{5\alpha}{3}$\,. This way, the volume scalar $\varphi$ is canonically normalised and the 7D metric is still in the Einstein frame. We leave $ds^2_3$ general for now (but with unit volume). We require only that it be Ricci flat. The $H_3$ flux is
\begin{equation}
H_3  = k\epsilon_3\,,
\end{equation}
where $k$ is a number. Note that $\epsilon_3$ here is the normalised volume element, so its integral gives 1. We will sometimes write vol$_3$ for the 3D volume form whose integral gives the volume. 

There is no $F_2$ flux since it is projected out so that the only possible contribution of its type would be a ``backreaction'' $F_2$ which vanishes in the smeared limit and in the EFT. The effective potential can be computed easily by dimensionally reducing the $H_3^2$, $F_0^2$ bulk terms in the 10D action together with the DBI term for the O6 planes. If we use the fact that the total O6 tension is minus the absolute value of the total O6 charge together with the RR tadpole:
\begin{equation}
dF_2 = F_0H_3 + \delta\quad \rightarrow \quad Q_6 = kM\,,
\end{equation}
we arrive at
\begin{equation}
S_7 =  \int \sqrt{|g_7|}\left[R_7 -\tfrac{1}{2}(\partial\varphi)^2 -\tfrac{1}{2}(\partial\phi)^2 - V(\phi, \varphi)\right]\,,
\end{equation}
where the potential takes the expected total square form: 
\begin{equation}
V(\phi, \varphi) = \frac{1}{2}\exp(\tfrac{3}{4}\phi + 7\alpha\varphi )\left(ke^{-\tfrac{7}{8}\phi + \tfrac{5\alpha}{2}\varphi} - F_0 e^{+\tfrac{7}{8}\phi - \tfrac{5\alpha}{2}\varphi} \right)^2 \,.
\end{equation}
The two exponentials inside the bracket are each others inverse and the scalar combination outside the bracket is exactly orthogonal to the scalar inside the bracket. This is the magic of no-scale structures. From the potential we deduce that the ``no-scale'' Minkowski vacuum lives at
\begin{equation}
e^{-\tfrac{7}{4}\phi + 5\alpha\varphi}  = \frac{F_0}{k}\,.
\end{equation}
We can rewrite this as
\begin{equation}\label{K/M}
 g_s Vol = \frac{k}{F_0}\,
\end{equation}
where $Vol$ is now the volume in 10D string frame. This shows that the weakly coupled limit $g_s\rightarrow 0$ implies a large volume limit $Vol\rightarrow \infty$, since the fraction $k/F_0$ is a fixed number. Hence, one concludes that the solution is perfectly tuneable in the supergravity regime. 

We end this section with a few more technical details about this background. An explicit realisation can be found by compactifying over $\mathbb{T}_3/\mathbb{Z}_2$, which gives 8 O6 fixed points. The resulting half maximal SUGRA in 7D was briefly described in \cite{Blaback:2015gwa}. The vacuum breaks supersymmetry completely and can be understood from an explicit T-duality of SUSY-breaking 4D Minkowski vacua in IIB, with 3-form fluxes and O3 planes \cite{Blaback:2015gwa} (see also \cite{Legramandi:2019ulq}). However, SUSY solutions are possible upon further breaking the Minkowski Poincar\'e symmetries by letting scalars run in a space-like direction, creating SUSY domain wall flows \cite{Blaback:2013taa, Blaback:2015gwa}. This gives a new interpretation to some of the SUSY massive IIA solutions that appeared in \cite{Janssen:1999sa, Imamura:2001cr}. Interestingly, these SUSY solutions can be constructed explicitly and lift to O6-NS5-D8 intersections.  For both localised and smeared O6 planes, crucial aspects like domain wall tension is left unchanged \cite{Blaback:2015gwa}. 

\section{Smearing and localising orientifolds}
In what follows, we present the \emph{general} forms of the smeared and localised O6 solutions. This is not new and can be found in \cite{Blaback:2010sj}, but we add them here to make this note self contained. We furthermore provide a general discussion about possible misconceptions concerning the smearing of orientifolds that we have encountered in papers and discussions. 

We start with the general form of the localised solution and then smear it. At the level of our discussion we do not need to specify any detail of the internal geometry, apart from it being conformal to a Ricci flat (compact) space.

The general solution in the 10D string frame reads: 
\begin{align}
\d s^2 &= S^{-1/2} \d s^2_7 + S^{1/2}\d s_3^2~,\nonumber\\
F_2 &= -g_s^{-1}\tilde{\star}_3 \d S\,,\nonumber\\
e^{\phi} &= g_s S^{-3/4}\,,\nonumber\\ \label{BPSbackground}
H_3 &=   F_0g_s \vol_3\,.
\end{align}
We denote the line element of 7D Minkowski space by
$\d s^2_7$ and the Ricci flat part of the internal space as $\d s_3^2$.
Here, $\tilde{\star}_3$ is the unwarped Hodge star derived from the metric $\d s_3^2$. 
The fluxes obey a Hodge-duality relation: 
\begin{equation}\label{ISD}
H_3 = \e^{\phi} \star_3 F_0~,
\end{equation}
where the Hodge-star $\star$ includes warp factors. This condition is identical to the famous \emph{Imaginary Self-Dual (ISD)} condition for ``fractional D3'' brane backgrounds, after performing three T-dualities along the Minkowski directions. One can furthermore verify that (\ref{ISD}) ensures a no-force condition for probe D6 branes. 

The warp factor function $S$ obeys a generalised Laplace equation that can be found for instance by solving the Bianchi identity for $F_2$:
\begin{equation}\label{BPS2}
\tilde{\triangle}_3 S +  \left(F_0 g_s\right)^2  = 4\pi g_s\sum_i\delta_i~,
\end{equation}
where the various $\delta$ sources correspond to the O6 positions and are individually normalised so that they integrate to unity.
Once this equation is solved, the ansatze (\ref{BPSbackground}) obey all 10D equations of motion. 

The difference between the smeared and localised solutions is straightforward: smearing corresponds to the limit $S\rightarrow 1$, which clearly is only feasible if the delta function in the Einstein, dilaton and Bianchi equations is smeared to a constant. But since the Einstein equation, dilaton equation and Bianchi identity all reduce to (\ref{BPS2}), this simply means that $4\pi g_s\delta \rightarrow (g_sF_0)^2$. This whole procedure is easily understood in a toroidal compactification where it would literally correspond to keeping all zero modes of all 10D supergravity equations and throwing away all higher Fourier modes. But since O-planes are localised objects which, unlike D-branes, cannot be stacked together in such a way as to produce a smearing effect, this procedure has been criticised. This criticism is not always justified however and the discussion is much more subtle.  ``Smearing'' rather means solving for the zero modes by truncating higher modes. This is obviously not correct since higher modes are typically sourced by zero modes. This is usually referred to as the issue of a ``consistent truncation'' \cite{Duff:1986hr}, which means zero modes (light modes) are not sources of higher (heavy) modes. Then a solution to the theory of the zero modes lifts to a full solution in 10D supergravity. Consistent truncations are not at all relevant in phenomenology where \emph{effective} field theories are constructed rather than exact truncations. But it is not always clear whether the effective field theory equals the theory without higher modes. In cases where supersymmetry restricts the lower-dimensional theory to unique couplings, we can check this explicitly. But most of phenomenology is based on minimal SUSY which in itself does not restrict the model enough to fully establish whether the effective field theory does in fact coincide with keeping only the zero modes. 

Warping in this context is an example of exciting massive modes since warping is sourced by the higher KK modes of the orientifold. Interestingly, there is a set of converging ideas on how to include warping in the EFT, see for instance \cite{Giddings:2005ff, Douglas:2008jx, Shiu:2008ry, Martucci:2014ska, Frey:2013bha, Frey:2008xw}. But one could ask a deeper question in the context of the scale separated vacua: \emph{does the existence of a $4D$ $N=1$ supergravity vacuum, coming from the zero modes, correctly informs us about the existence of a true higher-dimensional solution to the equations of motion with localised sources?} We will not answer this question but merely add that it can be thought of as fixing boundary conditions to second order PDE's and then hoping that the solutions have physical singularities. 

In any case, in the situation at hand, we will provide two examples in which we can solve the PDE's and verify the singularities are physical. In the second example, we will learn about a surprising aspect of how a space with a boundary can become boundary-less due to this backreaction. The reason we can achieve these explicit solutions is because the sources are not intersecting. Solving PDE's with intersecting localised sources, or even finding sufficiently concrete ansatze to get explicit PDE's, is in fact the major barrier to understanding whether the AdS scale separated vacua are truly there or not.

Before we go into the details, we address one source of criticism opposing the smearing of sources that has appeared in \cite{McOrist:2012yc, Douglas:2010rt}. The statement is that smearing orientifolds provides negative delocalised energy momentum inside the internal dimensions. Such energy-momentum tensors could then support backgrounds that fail to exist once the orientifolds are localised. For instance, consider a combination of a Romans mass and O6-planes in a compactification with Ricci flat internal dimensions (like CY). The Romans mass contributes the term
\begin{align}
g_s^{5/2} F_0^2\,,
\end{align}
to the energy momentum tensor. Hence, there is no dilution in the large volume limit, as compared to contributions from fluxes of higher forms which include inverse metric factors. So, the only way a CY geometry makes sense is if this energy density is compensated everywhere by the negative O6 energy. The CY limit appears not to be sensible when O6-planes are localised. 

We disagree with this logic and our torus example shows indeed that this logic is inconsistent. First of all, it is clear that the warp factors and conformal factors introduce delocalised terms in the Einstein equation that clearly compensate for the absence of smeared sources. Still, in a large volume limit, one would want the warping to be mild in order to trust the unwarped EFT. This works out rather straightforwardly. One can check in our example, and in the examples by De Wolfe et al \cite{DeWolfe:2005uu}, that the large volume limit is accompanied by a small $g_s$ limit precisely in such a way as to dilute each flux contribution identically. For instance $F_0^2$ might have no inverse volumes, but it has the highest coupling dependence, namely $g_s^{5/2}$. Hence, the large volume and weak coupling limits taken in \cite{DeWolfe:2005uu} and other examples do indeed approach the CY geometry. This is not proof that the localised solutions exist, but it certainly refutes the most straightforward objections.

\section{Examples}
We first recall our conventions for flux and charge quantisations. These are
\begin{align}
& \int F_p = (2\pi)^{p-1}(\alpha')^{(p-1)/2}f_p\,,\qquad f_p\in \mathbb{Z}\,,\\
& \d F_{8-p} =\ldots + (2\pi)^7\alpha'^4Q_p \delta_{9-p}\,.
\end{align}
The charge of a single Dp-brane, denoted $Q_p$ is
\be
Q_p = (2\pi)^{-p}(\alpha')^{-(p+1)/2}
\ee
and the charge of an Op-plane is $-2^{\,p-5}$ that of a Dp-brane. So an O6 plane has the charge of 2 anti-D6 branes. If the charge quantum of the Romans mass is $M$ and that of the $H_3$-flux is $K$, then the integrated form of the $F_2$ Bianchi identity is the RR tadpole condition
\be \label{tadpole}
KM = 2N_{O6} - N_{D6}\,,
\ee
where $N_{O6}$ and $N_{D6}$ denote the number of O6-planes and D6-branes respectively. Note that the relation (\ref{K/M}) in terms of quantised numbers becomes:
\be\label{quantisednumbers}
g_s Vol =(2\pi)^3 \frac{K}{M}\,.
\ee
Here, we expressed the volume in string units defined such that $\alpha'=1$. One can verify that this is consistent with the specific normalisation of coefficient in front of the delta functions in equation (\ref{BPS2}). 

\subsection{The $\mathbb{T}^3/\mathbb{Z}_2$ orientifold }

%%%%%%%%%%%%%%%%%%%%%%%%%%%%%%%%%%STEPH EDIT%%%%%%%%%%%%%%%%%%%%%%%%%%%%%%%%%%

The rectangular torus of {\it unwarped} volume $Vol=(2\pi L)^3$ has a line element
\begin{align}
ds^2_{\Tthree}=\left(2\pi L\right)^2\delta_{mn}d\sigma^m d\sigma^n \,, \qquad \forall \;\sigma\in \left[-\frac{1}{2},\frac{1}{2}\right]\,.
\end{align}
Here $\sigma$ is dimensionless. The torus is orientifolded by O6-planes defined through worldsheet parity involution multiplied by $(-1)^F$ and the associated target space $\mathbb{Z}_2$ involution:
\be
\sigma^n\rightarrow -\sigma^n\,.
\ee
There are 8 O6-planes at the fixed points $(\sigma^1,\sigma^2,\sigma^3)=(a,b,c)$ where $a, b, c$ can take the values $\{0, 1/2\}$. Instead of cancelling their RR tadpoles with 16 D6-branes, we will use fluxes. According to equation (\ref{tadpole}), we must have $KM=16$.

The solution is described in full detail by solving equation (\ref{BPS2}). To do this, one could attempt a Fourier decomposition of $S$ and plug this into the differential equation (\ref{BPS2}). But it is well known that this Fourier series does not converge. One therefore uses a regularised series, which is expressed in terms of Jacobi theta functions as explained in for instance \cite{Andriot:2019hay}. This is the approach we will follow as well. The $S$ function is then related to the notion of generalised Green's functions $G$ on compact spaces that obey
\be
(2\pi L)\Delta G(\bar{\sigma}) = \delta(\bar{\sigma})-\frac{1}{V} \,,
\ee
where $\bar{\sigma}$ is shorthand for the position vector, $\Delta$ is the Laplacian in dimensionless coordinates $\sigma^i$ and $V$ is the corresponding dimensionless volume.
The $1/V$ term ensures that the equation integrates to zero on both the right and the left hand side.  The general solution for the S-function is then given by
\be\label{Green}
S = \sum_{i=1}^8 q_i G(\bar{\sigma}-\bar{\sigma}_i) + S_0\,,
\ee
where $S_0$ is some constant, $i$ labels the various O6-planes at positions $\sigma_i$ and the $q_i$ are related to the O6 charges. Since we have a single O6 (and no D6) at every fixed point, we take
\be
q_i = 4\pi g_s\,,
\ee 
in string units ($\alpha'=1$)\footnote{Note that $32\pi = F_0^2 g_s V$.}.
The value of $S_0$ has to be such that the singularities in the metric are of the right kind to be interpreted as O6-planes\footnote{We refer to \cite{Cordova:2019cvf} for a clear exposition on orientifold singularities.} and we will fix it below. 
The regularised Green's function is given by
\begin{align}
G(\bar{\sigma})=\frac{1}{2\pi L}&\int_0^{\infty}dt\left(1-\prod_{j=1}^3\theta_3\left(\sigma^j|4\pi i t \right) \right)\,,\\
\theta_3\left(\sigma|4\pi i t \right)&=\sum_{n\in \mathbb{Z}}e^{2\pi i \left( n\sigma+2\pi i n^2 t \right)}=1+2\sum_{n=1}^{\infty}e^{-4\pi^2n^2t} \cos(2n\pi \sigma)\,,
\end{align}
where $\theta_3$ denotes the Jacobi theta function.\footnote{This function obeys the following property:\newline $\lim_{t\rightarrow 0^+}\theta_3\left(\sigma|4\pi i t \right)=\text{III}(\sigma)$, where $\text{III}(\sigma)=\sum_{n=-\infty}^{+\infty}\delta(\sigma-n)$ is called the Dirac comb.} In the fundamental domain $\sigma^i\in[-1/2,1/2]$, this represents a single delta function, at $\sigma=0$ and $t=0$, generated by the zero mode $n=0$. As $t\rightarrow 0^+$, all other contributions to $\theta_3$ at $\sigma\neq 0$ are strongly suppressed and $\theta_3(\sigma\neq 0|0)\rightarrow 0$ by definition of the delta function.
Taking into account all 3 dimensions, this corresponds to a delta function at $(\sigma^1,\sigma^2,\sigma^3)=(0,0,0)$ at the location of an O6 source.

To fix $S_0$, we have to zoom in near the singularity. It can be shown that, in this region, the Green's function approximates to \cite{Andriot:2019hay}
\be\label{approx}
G(\bar{\sigma}) = -\frac{1}{4\pi}\,\frac{1}{2\pi L}\,\frac{1}{|\bar{\sigma}|} + \ldots  
\ee
where the $\ldots$ denotes sub-leading behaviour. If we define $\bar{y}=2\pi L\bar{\sigma}$ the torus metric in $\bar{y}$ coordinates is identical to the metric on $\mathbb{R}^3$ in Cartesian coordinates and we find $G= -(4\pi|\bar{y}|)^{-1} + \ldots$.  This equation is enough to fix $S_0$, in analogy with O6-planes in $\mathbb{R}^3$. We find $S_0 = 1$ so that the approximation near the O6 source is
\be
S = 1 - \frac{g_s}{|\bar y|}\,,
\ee
in string units, as in flat space. Note that the $S$ function is positive but has a negative double derivative away from sources. Hence, it reaches a maximum in between O6-planes and becomes singular, diverging towards negative infinity, near the sources. The warp factor ($S$-function) is plotted for various values of $g_s/ L$ in Figure (\ref{WarpFact}) on a $\sigma^2=\sigma^3=0$ hyper-slice. It shows clearly how a decrease in $g_s/ L$ forces $S\sim 1$ over most of the internal manifold. It diverges to negative infinity very quickly close to the O6 source at the origin and at $\sigma^1= 1/2\sim \sigma^1=-1/2$. 
\begin{figure}[h]
	\centering
	\includegraphics[scale=0.7]{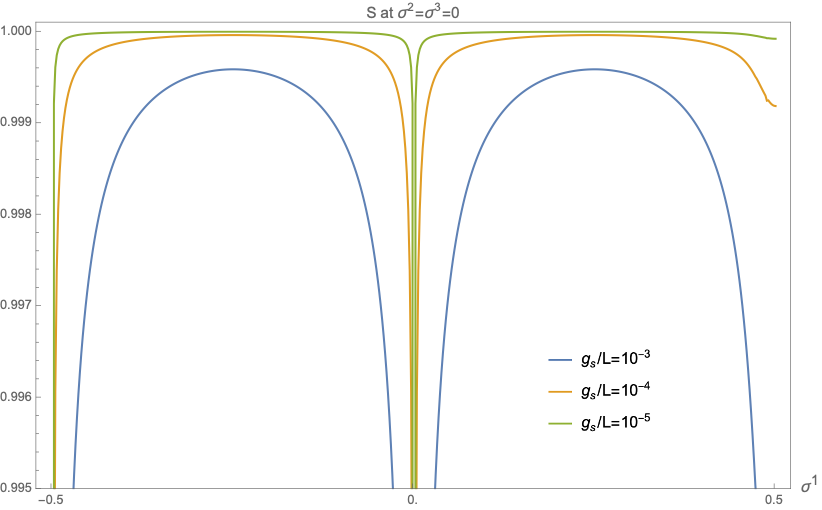}
	\caption{\small \emph{This figure shows the warp factor for various values of $g_s/ L$ on a $\sigma^2=\sigma^3=0$ hyper-slice. The orientifold causes it to diverge to negative infinity close to the source. For a small enough $g_s/ L$,  $S\sim 1$ over most of the internal space, consistent with the smeared approximation.}}\label{WarpFact}
\end{figure}
Note that the warp factor is only physically valid for $S>0$, where the metric is real and non-singular. $S$ becomes negative at a distance $|\bar\sigma_c|$ from the O6 sources, where the supergravity approximation ceases to be valid. 

The Green's function is a function of cosine's which is finite (the sum over modes converges for all $\sigma$) over the region $t>0$, while it is a simple delta function on the $t=0$ boundary. The expectation value of the warp factor over the volume of the torus is therefore
\begin{align}
	\langle S\rangle =\frac{1}{V} \int S(\sigma) &\;\text{vol}_3 =S_0+ \frac{1}{V}\sum_i q_i \int   G(\sigma-\sigma_i) \;d^3\sigma =S_0=1\,,
\end{align}
consistent with the smeared limit.

The full form of the coupling $e^{\phi}=e^{\phi_0}S^{-3/4}$ in terms of Jacobi theta functions is
\be\label{oupling1}
	e^{\phi}=e^{\phi_0}\left[1+2\frac{1}{L}g_s\sum_{i=1}^8\int_0^{\infty}dt\left(1- \prod_{m=1}^3\theta_3\left(\sigma^m-\sigma_i^m|4\pi i t \right)\right)\right]^{-3/4}
\ee
where the sum over the index $i$ represents a sum over O6 loci $\bar\sigma_i$.
Plots of the string coupling on the torus for various $|\bar\sigma_c|$ are shown in Figure \ref{stringcoup} on a $\sigma^2=\sigma^3=0$ hyper-slice. There are O6 sources at the origin and at $\sigma^1=1/2\sim \sigma^1=-1/2$. From this, it is clearly visible that an increase in volume forces the string coupling to $e^{\phi}=g_s$ in between the O6 sources as backreaction becomes quickly very weak away from the sources. 
\begin{figure}[h]
	\centering
	\includegraphics[scale=0.7]{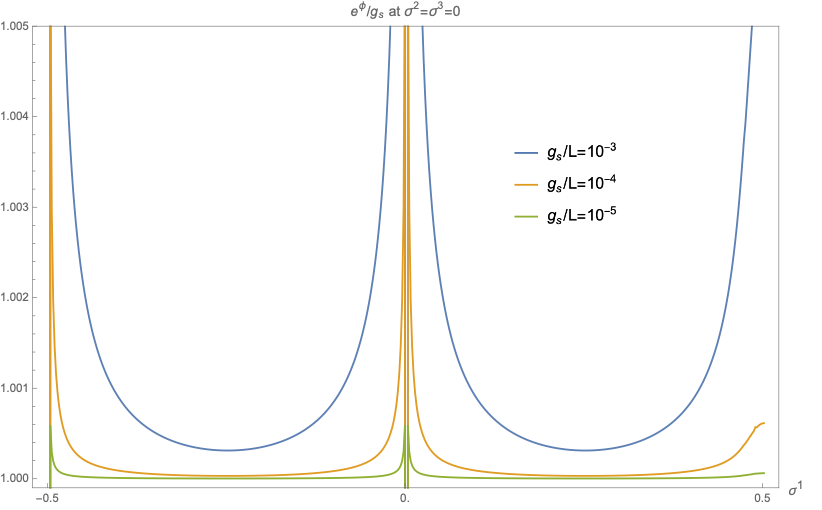}
	\caption{\small \emph{The dilaton $e^{\phi}$ is close to the vev predicted by the EFT, which we denote $g_s$, until it approaches the orientifold at the origin as seen on this $\sigma^2=\sigma^3=0$ hyper-slice. It diverges when $S=0$.}}\label{stringcoup}
\end{figure}

We have therefore verified explicitly that the smeared solution is a trustworthy approximation in the large volume and weak coupling limit in the sense that the localised solution quickly asymptotes to the smeared solution away from the sources. The numerics clearly shows this and the analytic argument is simple: the S-function in terms of the Jacobi-theta functions is written explicitly inside the brackets of the LHS of equation (\ref{oupling1}). There one sees that the higher modes, ie the extra contributions due to localisation, are suppressed by a $g_s/L$ factor. At the same time, $g_s/L$ is exactly the ``no-scale'' direction and can be made arbitrarily small. So in the limit of small $g_s/L$ one is bound to find agreement with the smeared solution.

\subsection{An exotic example}\label{sec:exotic}
In what follows we consider the space perpendicular to the 7D Minkowski space to be flat $\mathbb{R}^3$ in radial coordinates:
\begin{equation}
\d s_3^2= \d r^2 + r^2\d\Omega_2^2\,,
\end{equation}
and look for a solution with spherical symmetry. Since the internal space is now naively non-compact we will not insist on the tadpole immediately. The easiest solution one can find to (\ref{BPS2}) (without explicit $\delta$ sources) is:
\begin{equation}
S =   \frac{(F_0g_s)^2(r_0^2-r^2)}{6} ~,
\end{equation}
with $r_0$ some constant. The solution has a peculiar singularity for $r\to r_0$ and for the rest no obvious explicit D6 or O6 sources in it. In order to examine the local behaviour of the singularity we expand around it via $r=r_0-\delta r$, to find 
\be\label{metric1}
ds^2 = \sqrt{\frac{\alpha}{\delta r}} ds_7^2 + \sqrt{\frac{\delta r}{\alpha}}\left( (d\delta r)^2 + r_0^2\d\Omega^2\right)\,,
\ee
where $\alpha^{-1}= r_0 (F_0g_s)^2/3$. Note that $ds_7^2$ is flat Minkowski space and any constant rescaling of that metric can be gauged away. Let us compare this with an  O6-plane singularity. The metric of an O6-plane (in flat space) is given by
\be
ds^2 = \frac{1}{\sqrt{H}} \d s^2_7 + \sqrt{H}\left( d x^2 + x^2 d\Omega^2\right)\,,
\ee
where $H =  1-\frac{g_s}{x}$. The singularity occurs at $x=g_s$. When expanding the metric to first order via
$x =  g_s +\delta x$, we find:
\be\label{metric2}
ds^2 = \sqrt{\frac{g_s}{\delta x}} ds_7^2 + \sqrt{\frac{\delta x}{g_s}} (d\delta x)^2 + \sqrt{\delta x}g_s^{3/2}\d\Omega^2\,.
\ee
We could now attempt to match the two metrics (\ref{metric1}, \ref{metric2}) by taking $\delta x$ and $\delta r$ proportional to each other:
\be
\delta r=\beta\delta x\,.
\ee
Then (\ref{metric1}) becomes
\be
ds^2= \sqrt{\frac{\alpha}{\beta \delta x}} ds_7^2 + \beta^{5/2}\sqrt{\frac{\delta x}{\alpha}} (d\delta x)^2 + r_0^2\sqrt{\frac{\beta}{\alpha}}\sqrt{\delta x}\d\Omega^2\,.
\ee
We find that we can match the singularities if
\be
 \beta^{5/2}=\alpha^{1/2}g_s^{-1/2}\,,\qquad  F_0^2r_0^6=3g_s^2\,.
\ee
We can rewrite the second condition, in terms of flux quanta as
\be\label{prize}
r_0^3 =\frac{g_s}{|M|}2\pi\sqrt3\,.
\ee
So at the cost of having to fix $r_0$ in a substringy regime (\ref{prize}) we see that the local structure of the singularity of our flux background has exactly the same local form as the near-singularity expansion of an O6-plane. This can be confusing at first since $r=r_0$ is the 2-sphere boundary of a 3-ball, but in the warped space it truly corresponds to a localised O6 singularity. To really be sure the whole 3d space with (orientifold) singularity can be considered compact we should check the tadpole condition and flux quantisation. 

For a single O6-plane we have $KM=2$ and both $|M|$ and $|K|$ are bounded by 2. We furthermore have (\ref{quantisednumbers}) where $Vol$ in that formula  corresponds to the \emph{unwarped} manifold, which in this case is a ball of radius $r_0$ where $ Vol = 4\pi r_0^3/3$.\footnote{Note that the warped volume is
\begin{equation}
{\cal V} = \f{\pi^2 r_0^2\sqrt{6}}{F_0g_s}~.
\end{equation}}
This does show that the tadpole condition is automatically satisfied, but we need $KM=2$. We therefore obtained a compact background starting from an unwarped space that is really a ball, a space with a boundary. And we found it was consistent to interpret the boundary of the ball as the location of an O6-plane, despite the boundary, an $S^2$, being co-dimension 1 instead of the required 3. The reason this works is that the backreaction curves the space such that the $S^2$ shrinks to stringy size and becomes consistent with the singularity induced by a localised co-dimension 3 O6-plane. To make this intuitively more clear we have presented a 2D analogue of this in figure \ref{Figure}.

\begin{figure}[ht!]
	\begin{center}
		\includegraphics[scale=0.25]{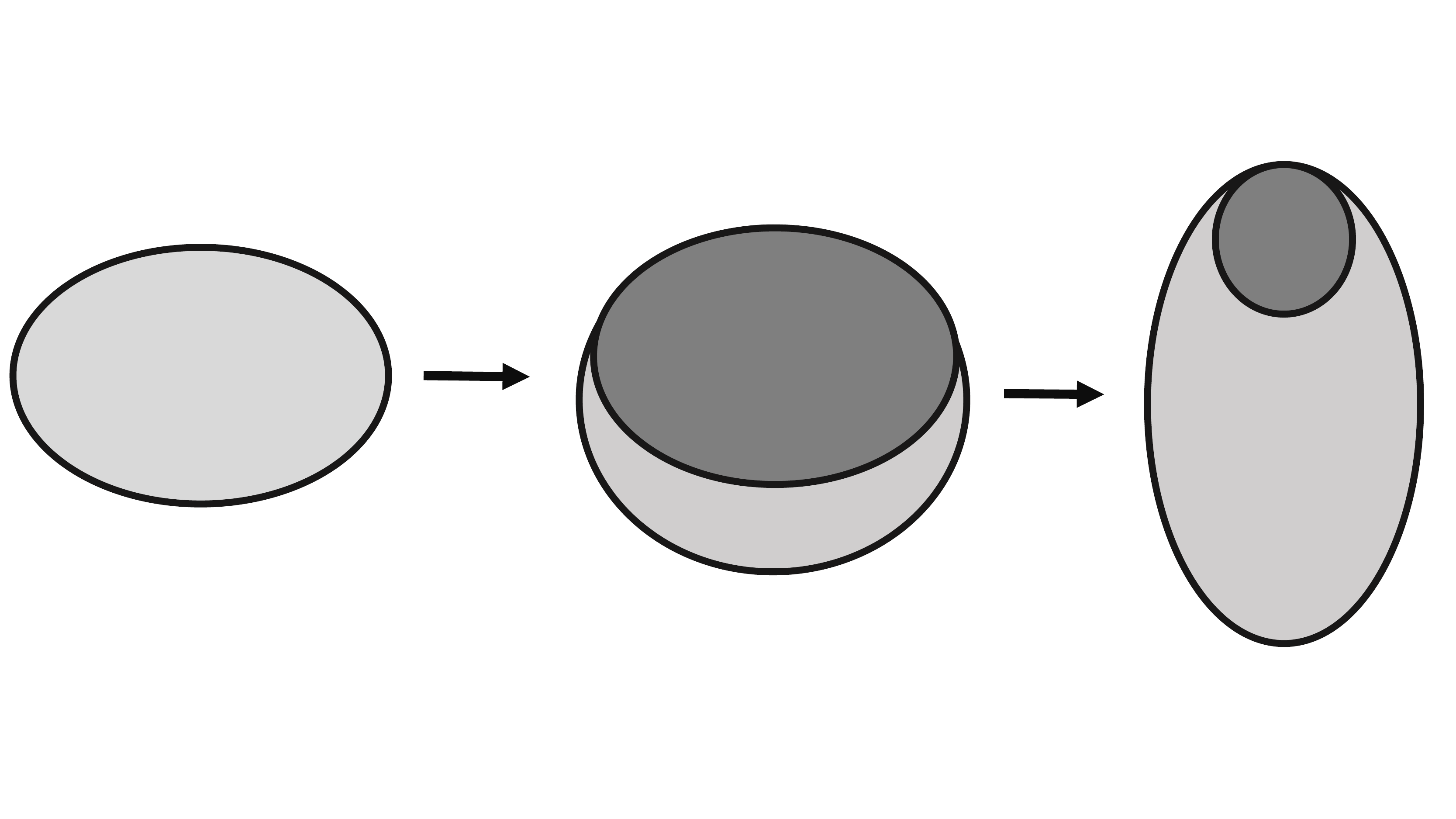}
		\caption{\it A disk gradually loosing the boundary, by turning into a singular $S^2$.}
		\label{Figure}
	\end{center}
\end{figure}
Note that usually orientifolds are described in a covering space by first specifying a $\mathbb{Z}_2$ action. Here we got immediately the resulting space. So the full covering space is just designed by attaching two such ``droplet'' geometries to each other near the O6 so that they mirror each other through the O6. 

We see two major qualitative differences with the previous orientifold solution on the torus: 1) the volume is fixed in terms of the coupling due to equation (\ref{prize}). There is no modulus here and so the solution is not described by our EFT we had earlier. 2) since we cannot take a large volume and weak coupling limit at the same time, the solution is not at all in the supergravity regime and is only a formal solution to the supergravity equations of motion, but its stringy meaning is completely unclear. 

In what follows we speculate on why we find a clash with the EFT description.

\subsection{Comments on the EFT}

In order for flux solutions to make sense one has to make sure one can trust the approximations that were used to construct them. We made use of 10D supergravity with localised sources as an approximation to IIA superstring theory. A minimal requirement therefore is that the string coupling constant $e^{\phi}$, the curvature scales, and inverse volumes of compact cycles are all small. If not, string loop corrections, derivative corrections or corrections from stringy degrees of freedom that become light will dominate. Whenever sources are included we cannot insist on some of these properties since the supergravity description of D-branes and orientifolds is singular to start with. Therefore the usual attitude is that the region near the sources should be singular in the same way as they would be when inserted in 10D flat space where we have a worldsheet understanding of them. We furthermore  require that the region outside of the sources, where their backreaction should become negligible, obeys all our above criteria of large volumes, weak coupling and small curvature. Whether this attitude is sufficiently safe is unclear. For instance in the presence of Romans mass we cannot lift O6-planes to 11D to understand their strong coupling behaviour from S-duality with M-theory \cite{Seiberg:1996nz}.  

Our 7D EFT analysis obtained from integrating the action over the extra dimensions informed us, through equation (\ref{K/M}), that the string coupling and internal volume can both be put arbitrarily well in their respective corners where we should trust our approximations. We verified this explicitly by solving the $\mathbb{T}_3/\mathbb{Z}_2$ example where the fields indeed quickly asymptote to the values predicted by the EFT away from the sources, such that most of the internal manifold is weakly coupled and curved. This is expected in general for a simple reason. Consider Fourier expanding the coupling over the 3-torus:
\begin{equation}
e^{\phi} = \sum_{\vec{n}} c_{\vec{n}} \exp(2\pi i \vec{n}\cdot\vec{\theta})\,.
\end{equation}
Then the integral over the internal space clearly picks out the zero mode. Hence the average value of the coupling, or any other field, will be given by its zero mode. If the average is small, so should the field be over the whole manifold. The only exception would be if the field fluctuated very strongly to high values away from sources while still keeping the average low. There is no real reason why that should or could happen in our opinion. Of course the mode expansion tends to not converge and the use of Jacobi-theta functions in our explicit analysis made clear one has to use regularised expressions. Nonetheless we expect the spirit of the above discussion to remain correct and our numerics did confirm this.

We explained that our exotic solution discussed in section \ref{sec:exotic} does not obey the EFT expectations at all. So we conclude that nowhere the solution is in the regime of 10D supergravity.
There is a simple explanation for why this happens: Consider some compactification on a manifold that has a boundary, prior to backreaction of the sources. Then we expand some field, denoted $\Phi$,  in eigenmodes of the Laplacian, denoted $\phi_n$:
\begin{equation}
\Phi = \sum_{n} c_n \phi_n
\end{equation}
where we assumed a countable spectrum labeled by an integer $n$:
\begin{equation}
\Delta \phi_n =\lambda_n\phi_n\,,\quad \lambda_0=0\,,\quad \lambda_n\neq 0 \,\,\text{if}\,\,n\neq 0.
\end{equation}
Due to Stokes theorem we have that for spaces \emph{without} boundaries:
\begin{equation}
\int \Phi = \phi_0\,,
\end{equation}
and hence the zero modes inform us about averages. This relation does not hold on spaces with boundaries, even if the boundaries disappear after properly backreacting the sources. This explains why our exotic example fails to satisfy the weak coupling and curvature conditions.

\section{Conclusion}
We have provided a proof of principle that O6-planes in flux compactifications with Romans mass can lead to well behaved solutions, at the 10D supergravity level, by explicitly working out a compactification to 7D Minkowski space on $\mathbb{T}_3/\mathbb{Z}_2$ and solving explicitly for all fields. We found that the solution can indeed be arbitrarily weakly coupled and curved away from the O6 sources as predicted by the 7D supergravity which captures the smeared orientifold solution as we emphasised. The limit of weak coupling and large volume is due to a flat direction, which is expected to be lifted in the full string theory. Hence the fate of this background in full string theory remains not well understood, but some more down-to-earth worries about the supergravity solutions being weakly coupled and curved away from sources may be put to rest by this work.

This does not imply that the massive IIA reductions to scale separated AdS vacua of \cite{DeWolfe:2005uu, Derendinger:2004jn}  are certainly under control. Although recent progress in\cite{Junghans:2020acz, Marchesano:2020qvg} might point in that direction, it is still somewhat unclear at this point. And the issues raised in \cite{Banks:2006hg} are still there.  At least our constructions show that some of the most straightforward worries, like the use of massive IIA supergravity with O6 sources are not issues themselves. What complicates the AdS vacua is that the orientifolds intersect, whereas the examples in this paper all featured parallel sources. 

To our knowledge there are hardly any fully explicitly worked out compact flux solutions with backreacted orientifold sources and we hope this example will be more widely useful as well\footnote{Other examples can be found in \cite{Apruzzi:2017nck, Cordova:2019cvf} but there is an ongoing debate as to  whether the orientifold singularities are really physical \cite{Cribiori:2019clo,Cordova:2019cvf}.}. For instance our torus example explicitly shows that the ``wormhole like'' desingularisation of an O6-plane found in \cite{Saracco:2012wc} is unfortunately not a generic feature and as argued in \cite{Gautason:2015tig} is also not expected in the scale separated AdS vacua. 

The second result of our paper came from analysing an exotic orientifold solution (which was already discussed in passing in \cite{Danielsson:2016cit, Blaback:2019ucp}) with the property that the space on which the orientifold is inserted has a boundary (the space is a 3-ball). But after backreacting the orientifold the boundary pinches off into a co-dimension 3 source that effectively renders the manifold compact. To our knowledge this provides an entirely new kind of flux solution, which is difficult to find from an EFT approach. The reason is that the EFT does not capture the qualitative features of the solution as we explained. Technically this happens because the boundary makes the integral over the extra dimensions pick up extra boundary terms and not just the zero modes of fields.   Unfortunately this very feature also leads to the exotic solution being strongly coupled almost everywhere. So in that sense we have not found a new flux solution to string theory. %Although it could be that adding a few D6 branes would lower the coupling in a larger part of the manifold, but it remains very unclear whether it would be sufficiently weakly coupled on average. 

Since our solutions were obtained from T-duality of standard type IIB compactifications \cite{Dasgupta:1999ss,  Giddings:2001yu} with O3 sources (by smearing along 3 directions), we consider it very sensible that similarly, one can start considering flux compactifications of IIB down to 4D Minkowski spaces on 6D manifolds that have boundaries prior to backreaction. This would significantly enlarge the flux landscape. However, it is reasonable to expect that the solutions escape a proper weak coupling or large volume limit for reasons similar to our exotic O6 example.  

\section*{Acknowledgements}
We are very grateful to Fri{\dh}rik Gautason for his work during the initial stages of this project. This research was supported in part by the National Science Foundation under Grant No. NSF PHY-1748958. The work of TVR is supported by the KU Leuven C1 grant ZKD1118 C16/16/005.

{\small
\bibliographystyle{utphys}

    \bibliography{refs}
}

\end{document}